\input{aipcheck}

\documentclass[
    ,final            
  ]
  {aipproc}

\layoutstyle{6x9}

\usepackage{amssymb}
\usepackage[varg]{txfonts}

\def\aj{AJ}%
\def\araa{ARA\&A}%
\def\apj{ApJ}%
\def\apss{Ap\&SS}%
\def\aap{A\&A}%
\def\mnras{MNRAS}%
\def\prl{Phys.~Rev.~Lett.}%
\begin{document}

\title{Three-dimensional Modeling of Type Ia Supernova Explosions}


\author{F.~K.~R{\"o}pke}{
  address={Max-Planck-Institut f\"ur Astrophysik,
  Karl-Schwarzschild-Str.~1, D-85741 Garching, Germany}
}

\author{W.~Hillebrandt}{
  address={Max-Planck-Institut f\"ur Astrophysik,
  Karl-Schwarzschild-Str.~1, D-85741 Garching, Germany}
}

\begin{abstract}
Modeling type Ia supernova (SN Ia) explosions in three dimensions allows to
eliminate any undetermined parameters and provides predictive power to
simulations. This is necessary to improve the understanding of the
explosion mechanism and to settle the question of the applicability of
SNe~Ia in cosmological distance measurements. Since the models contain
no tunable parameters, it is also possible to directly assess their
validity on the basis of a comparison with observations. Here, we
describe the modeling of SNe Ia as thermonuclear
explosions in which the flame after ignition near the center of the
progenitor white dwarf star propagates outward in the sub-sonic
deflagration mode accelerated by the interaction with turbulence. We
explore the capabilities of this model by comparison with observations
and show in a preliminary approach, how such a model can be applied to
study the origin of the diversity of SNe Ia.
\end{abstract}

\maketitle


\section{Introduction}

Type Ia supernovae (SNe~Ia) have become one of the major tools in
observational cosmology. Yet a sound theoretical understanding of
these objects -- justifying in particular the calibration techniques
applied in distance measurements -- is still lacking.
First attempts to model SNe~Ia were based on one-dimensional numerical
simulations. Such models gave valuable insight into the basic
mechanism of SN~Ia explosions. However, their predictive power is
limited due to the fact that underlying physical processes enter the
models in a parametrized way. This is overcome by three-dimensional
modeling of SNe~Ia \citep{reinecke2002d,gamezo2003a}.
As an example, we present a model that is derived from the standard scenario of
SN~Ia explosions (for a review see \citep{hillebrandt2000a}). A white
dwarf (WD)
consisting of carbon and oxygen is assumed to accrete matter from a
non-degenerate binary companion until its mass approaches the
Chandrasekhar limit. Due to the rapid increase of the central density
nuclear reactions ignite giving rise to a stage of convective carbon
burning. This stage lasts for several hundred years and terminates
once the nuclear energy production cannot be balanced by convective
cooling any longer. Subsequently, a thermonuclear runaway of a small
temperature fluctuation ignites a thermonuclear flame. The exact
mechanism of flame ignition, however, remains controversial. While
some studies suggest a flame ignition in multiple sparks distributed
around the center of the WD
\citep{garcia1995a,woosley2004a,iapichino2005a,kuhlen2005a}, others
put forward central single-point ignitions \citep{hoeflich2002a}.

After ignition the flame propagation is determined by the laws
of hydrodynamics. Regarding the flame front as a discontinuity between
fuel and ashes, they allow for two distinct modes of flame
propagation. In a sub-sonic deflagration burning is
mediated by thermal conduction while a
super-sonic detonation is driven by a shock wave. On the basis of
one-dimensional simulations of a prompt detonation,
\citet{Arnett1969a} ruled out this model for SNe~Ia since it
drastically underproduces intermediate mass elements observed in
spectra of these events.

Starting out as a laminar deflagration, however, the flame propagates too
slowly to explain the energy release necessary to explode the
WD. Thus, any valid SN~Ia model needs to provide means of flame
acceleration. Two mechanisms are conceivable here. Firstly, the flame
propagation may continue in the deflagration mode being significantly
accelerated by the interaction with turbulence. 
The one-dimensional model \emph{W7} of \citet{nomoto1984a}
demonstrated that such a model is in principle capable of reproducing
the main observational features of SNe~Ia. An alternative way to
speed up the flame is to assume a deflagration-to-detonation
transition (DDT) at later stages of the explosion. The weak point of
these delayed detonation models (e.g.~\citep{gamezo2004a}) is that
a physical mechanism providing a DDT in SNe~Ia could not be identified yet
\citep{niemeyer1999a,lisewski2000b,roepke2004a,roepke2004b}
and therefore the hypothetical transition of the burning mode enters the model as an
undetermined parameter.

\section{A deflagration type $\mbox{Ia}$ supernovae model}

Our goal in the following is to present a SN~Ia explosion model that
contains no tunable parameters. Therefore we set aside the possibility
of a delayed detonation and focus on the turbulent deflagration model.
Turbulence is induced here by generic instabilities. The flame
propagates from the center of the star outward leaving behind light
and hot ashes. Dense and cold fuel in front of the flame
gives rise to an inverse density stratification in the gravitational
field of the WD. This renders the flame propagation buoyancy unstable
and in its nonlinear stage the Rayleigh-Taylor instability leads to
the formation of (typically mushroom-shaped) burning bubbles that rise
into the cold fuel. At the interfaces of these bubbles strong shear
flows emerge. The corresponding Reynolds numbers reach values of the
order of $10^{14}$ and therefore strong turbulence is generated by
secondary shear (Kelvin-Helmholtz) instabilities. The turbulent eddies
generated on scales of the buoyancy-induced flame features decay to
smaller scales forming a turbulent energy cascade and interact with the
flame propagation. This stretches and corrugates the flame enlarging
its surface area and thus the net burning rate is increased.

The main challenge in numerically implementing this scenario is the
vast range of relevant length scales involved. Not only is the width
of a thermonuclear flame in the degenerate carbon/oxygen material at the onset of the
explosion 9 orders of magnitude below the radius of the WD. The
turbulent cascade extends to even smaller scales and interacts with
the flame down to the Gibson length at which the laminar flame speed
equals the turbulent velocity fluctuations ($10^4\, \mathrm{cm}$ and decreasing
in the explosion process). This problem can be tackled
in a Large Eddy Simulation (LES) approach. Here, only the
largest scales of the problem are directly resolved (applying the
\textsc{Prometheus} implementation \citep{fryxell1989a} for solving the hydrodynamics
equations) and turbulence effects
on unresolved scales are included via a subgrid-scale model
\citep{niemeyer1995b,schmidt2005c}. The thermonuclear flame is modeled
as a sharp discontinuity separating the burnt from the unburnt
material. Its evolution is followed utilizing the level-set method
\citep{osher1988a}. Since the structure of the flame is not resolved,
the flame propagation velocity must be provided externally. This,
however, does not introduce an undetermined parameter to the model
since the theory of turbulent combustion \citep{peters2000a} predicts
that for most stages of the SN~Ia explosion the flame propagation
proceeds in the flamelet regime where it completely decouples from the
microphysics of the burning and is determined by the turbulent
velocity fluctuations that can be derived from the subgrid-scale
model. This implementation provides a self-consistent model of SN~Ia
explosions in the deflagration scenario.

Our description of burning is augmented by a simplified treatment of
the nuclear reactions including only five species
\citep{reinecke2002b}. It provides the energy release necessary to
follow the explosion dynamics. In order to derive observables from the
models, however, the chemical composition of the ejecta needs to be
known in detail. This is achieved in a nucleosynthesis postprocessing
step \citep{travaglio2004a}. Of particular interest is the yield of
$^{56}$Ni, since its radioactive decay powers the light curve.

\begin{figure}
  \centerline{\includegraphics[width=0.87\textwidth]{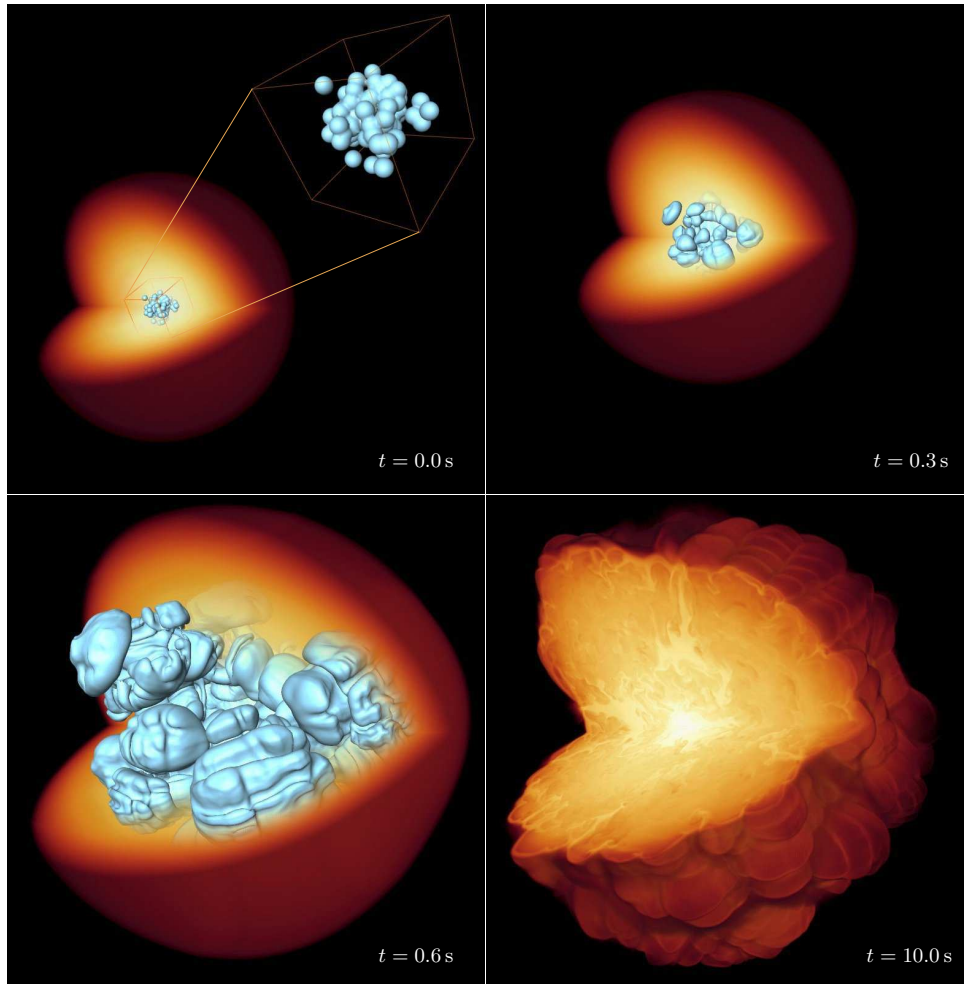}}
  \caption{Snapshots from a full-star SN~Ia simulation starting from
  a multi-spot ignition scenario. The logarithm of the density is
  volume rendered indicating the extend of the WD star and the
  isosurface corresponds to the thermonuclear flame. The last snapshot
  marks the end of the simulation and is not on scale with
  the earlier snapshots.\label{fig:evo}}
\end{figure}

\section{Simulation results}

A typical evolution of a SN Ia explosion modeled as described above is
shown in Fig.~\ref{fig:evo}. Starting from an ignition in multiple
sparks the flame propagates outward. At $t = 0.3 \, \mathrm{s}$, the
mushroom-shaped features due to the buoyancy instability are clearly
visible. Subsequently, the flame becomes increasingly corrugated and
is accelerated by interaction with turbulence. It therefore
burns through a large fraction of the WD material. The snapshot at $t
= 0.6 \, \mathrm{s}$ shows the flame evolution around the peak of
energy production due to nuclear burning. Up to this point, the burning
terminated in nuclear statistical equilibrium (NSE) and the
carbon/oxygen material was primarily converted to iron group
elements. The expansion of the WD decreases the fuel density
steadily and once it falls below $5 \times 10^7 \, \mathrm{g}\,
\mathrm{cm}^{-3}$ nuclear burning becomes incomplete and produces
mainly intermediate mass elements. About $2 \, \mathrm{s}$ after
ignition, expansion quenches the burning and the following evolution
is characterized by the relaxation to homologous expansion of the ejecta,
which is reached to a reasonable accuracy $\sim$$10 \, \mathrm{s}$
after ignition \citep{roepke2005c}. The density structure of the ejecta at this stage is
shown in Fig.~\ref{fig:evo}, where the traces of turbulent flame
propagation are clearly visible.

\begin{table}
\begin{tabular}{lrrr}
\hline
    \tablehead{1}{l}{b}{Parameter}
  & \tablehead{1}{r}{b}{Range of variation}
  & \tablehead{1}{r}{b}{Effect on $^{56}$Ni production}
  & \tablehead{1}{r}{b}{Effect on total energy}\\
\hline
      X($^{12}$C)  & [0.30, 0.62]  & $\le$$2\%$ & $\sim$14\% \\
      $\rho_\mathrm{c}$ [$10^9 \, \mathrm{g}/\mathrm{cm}^3$]  &
      [1.0, 2.6] & $\sim$6\% & $\sim$17\% \\
      $Z$ [$Z_\odot$] & [0.5, 3.0] & $\sim$20\% & none \\
\hline
\end{tabular}
\caption{Variation of initial parameters in SN~Ia explosion
      models}
\label{tab:div}
\end{table}

Apart from the initial conditions simulations as described above
contain no free parameters. Therefore the question arises whether
such models are capable of reproducing observations without any
fitting. The explosion energies achievable in the outlined scenario
reach up to $\sim$$8 \times 10^{50}\,\mathrm{erg}$ and the models
produce $\sim$$0.4 \, M_\odot$ of $^{56}$Ni. This falls into the range
of observational expectations, although on the side of the weaker
SN~Ia explosions \citep{contardo2000a,stritzinger2005a}.
Nonetheless, the synthetic lightcurves derived from models of the
class described here fit the observations in the $B$ and $V$ bands
around maximum luminosity rather well
\citep{sorokina2003a,blinnikov2005a}. 
A much harder constraint on the explosion model is posed by spectral
observations, since spectra are particularly sensitive to the chemical
composition of the ejecta. \citet{kozma2005a} pointed out a potential
problem of deflagration SN~Ia models. In late time ``nebular'' spectra,
unburnt material (transported towards the center in downdrafts due to
the large-scale buoyancy-unstable flame pattern) gives rise to a
strong oxygen line of low-velocity material which is in conflict with
observations. However, the synthetic spectrum of \citep{kozma2005a} was
derived from a simplistic centrally ignited model. Recently, 
\citet{roepke2005e} showed that multi-spot ignition models may succeed
to burn out the central parts of the WD, reducing the amount of oxygen
at low velocities. A stochastic multi-spot ignition leads to similar
results \citep{schmidt2006a}.
Detailed spectral observations allow to determine the chemical
composition of the ejecta in velocity space \citep{stehle2005a}. The
mixed composition of the ejecta observed there points to a deflagration
phase being at least a significant contribution to the SN~Ia explosion
process. The central parts are found to be clearly dominated by iron
group elements, which are mixed out to velocities of about $12000 \,
\mathrm{km} \, \mathrm{s}^{-1}$. Intermediate mass elements are distributed
over a wide range in radii and no unburnt material is found at
velocities $\lesssim$$5000  \, \mathrm{km} \, \mathrm{s}^{-1}$. A
recent high-resolution full-star deflagration SN~Ia simulation
\citep{roepke2005f} demonstrated that such models can get close to
these observational constraints.

Although it cannot be ruled out that the pure deflagration model of
SNe~Ia is incomplete, the results indicate that it may be at least a
dominant part of the mechanism. Therefore it is justified to ask how
such models are affected by initial parameters of the exploding
WD. This may give a hint to the origin of the diversity of SNe~Ia.
To moderate the computational expenses, simplified setups may be used
to study the effects of physical
parameters on the explosion models. Such an approach was recently
taken by \citet{roepke2005d} and resulted in the first systematic study of
progenitor parameters in three-dimensional models. The basis of this
study was a single-octant setup with moderate (yet numerically
converged) resolution. However, the lack in resolution did not allow for a
reasonable multi-spot ignition scenario and thus only weak explosions
can be expected. It is therefore not possible to set the absolute
scale of effects in this approach, but trends can clearly be
identified. 
The parameters chosen for the study were the WD's
carbon-to-oxygen ratio, its central density at ignition and its
$^{22}$Ne mass fraction resulting from the metallicity of the
progenitor. All parameters were varied independently to study the
individual effects on the explosion process. The results of this survey are given in
Tab.~\ref{tab:div}.
\begin{figure}
\centerline{
\includegraphics[width=0.87\textwidth]{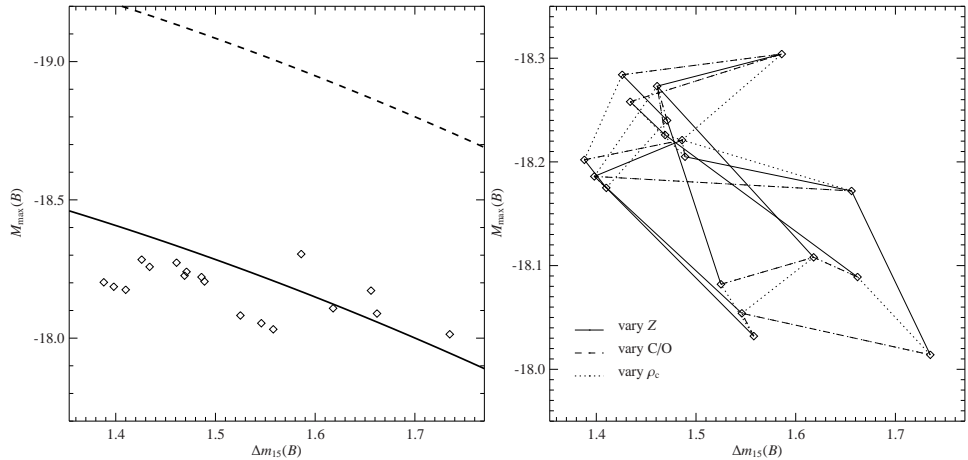}}
\caption{Peak luminosity vs.\ decline rate of the light curve in the B
  band (diamonds correspond to SN Ia explosion models; the dashed
  curve in the left panel indicates the
  original relation by \citet{phillips1999a}
  and a shifted relation is marked as a solid curve)\label{fig:ph}}
\end{figure}
To determine the effects of these variations on observables, synthetic
light curves were derived from all models \citep{blinnikov2005a}. From these, the peak
luminosities and decline rates (in magnitudes 15 days after maximum;
$\Delta m_{15}$) were determined.
A comparison with the relation given by
\citet{phillips1999a} (forming the basis of the calibration of
cosmological distance measurements) is provided in the left panel of
Fig.~\ref{fig:ph}. Obviously, the absolute magnitude of the
\emph{Phillips relation} is not met by our set of models. Moreover, the range
of scatter in $\Delta m_{15}$ is much narrower than that of the set of
observations used by \citet{phillips1999a}. But there is a trend of our models
consistent with the slope of the
Phillips relation. 
The right panel of Fig.~\ref{fig:ph} shows that this slope is
dominated by the variation of the progenitor's metallicity.


\begin{theacknowledgments}
F.K.R.~gratefully acknowledges the kind invitation to the OMEG05
conference in Tokyo and friutful discussions with K.~Nomoto and his
group on SN~Ia explosions.
\end{theacknowledgments}

\end{document}